\documentclass[a4paper,12pt]{article}
\usepackage{amsmath,amsfonts,amssymb,cite}
\usepackage{graphicx}

\newcommand{\be}{\begin{equation}}
\newcommand{\ee}{\end{equation}}

\begin{document}

\begin{center}
{\large\bf The case of equivalence of low and high energy constraints on Regge vector spectrum in AdS/QCD}
\end{center}
\bigskip
\begin{center}
{ S. S. Afonin
and T. D. Solomko}
\end{center}

\begin{center}
{\small\it Saint Petersburg State University, 7/9 Universitetskaya nab.,
St.Petersburg, 199034, Russia}
\end{center}

\bigskip

\begin{abstract}
The AdS/QCD models are believed to interpolate between low and high energy sectors of QCD.
This belief is usually based on observations that many phenomenologically reasonable predictions
follow from bounds imposed at high energies although the hypothetical range of applicability of
semiclassical bottom-up holographic models  is restricted by the gauge/gravity duality to low energies
where QCD is strongly coupled. For testing the feasibility of high energy constraints it is interesting to calculate
holographically some observable constants at low and high momenta independently and compare.
We consider an AdS/QCD model describing the Regge-like linear spectrum of spin-1 mesons in a general form
and show that under definite physical assumptions, the low-energy constraints on 2-point correlation
functions lead to nearly the same numerical values for the parameters of linear radial spectrum as the
high energy ones. The found approximate coincidence looks surprising in view of the fact that such a
property for observables is natural for conformal theories while real strong interactions are not conformal.
\end{abstract}

\section{Introduction}

The bottom-up holographic models for QCD (often referred to as AdS/QCD models) have proven to be an
interesting approach to the phenomenology
of strong interactions~\cite{br}. Being inspired by the gauge/gravity duality in string theory~\cite{mald,witten},
they boldly apply the holographic methods developed for conformal theories to the case of real QCD which
is not conformal. A theoretical justification for such an extension is still lacking (see, however,
the pioneering work~\cite{pol}), nevertheless
the bottom-up holographic approach resulted in construction of some useful phenomenological models
which turned out to be unexpectedly successful~\cite{son1,pom,son2} and triggered a large activity in the field
(the Refs.~\cite{bottom-up,bottom-up2,bottom-up3,bottom-up4,bottom-up5,bottom-up6,bottom-up7,bottom-up8,bottom-up9,bottom-up10,bottom-up11,bottom-up12,andreev,holSR,holSR2,No-wall,SWaxial,genSW,forkel,UV,UV2,UV3,nonlSW,holog2010,son3,zuo} constitute only a tiny part).

According to the principles of gauge/gravity duality, the AdS/QCD models should be viewed as models for QCD
constructed (i) in the large-$N_c$ limit~\cite{hoof,wit} and (ii) in the low energy domain where QCD is strongly coupled.
The point (i) formally implies that only 2-point correlation functions can be described self-consistently (the higher n-point functions
vanish in the large-$N_c$ limit~\cite{wit}) while the point (ii) ensures that a putative 5D dual gravitational theory
is weakly coupled, hence, can be treated semiclassically. In phenomenological applications, however, one often weakens
both requirements. For instance, the incorporation of Chern-Simons terms allows to describe various $\mathcal{O}(1/N_c)-$effects
of anomalies. Also the descriptions of spontaneous Chiral Symmetry Breaking (CSB)
and hadron formfactors involve higher n-point correlators~\cite{br,son1,pom}.
The applications to high energy QCD, which look speculative in view of (ii), turn out to be rather successful.
As an example one can mention holographic
derivations of QCD sum rules~\cite{holSR,No-wall,SWaxial,genSW}. Among the practitioners of AdS/QCD models there is
a widespread belief that these models efficiently interpolate
between low energy and high energy QCD. In certain sense, they can be seen as a "meromorphization" of the perturbative QCD
expression for 2-point correlation functions~\cite{br}. Since essentially the same is pursued in the large-$N_c$ QCD sum
rules~\cite{sr,sr2,sr3,sr4,sr5,sr6,sr7,sr8,sr9,we,npb,lin}, the holographic analogues of these sum rules provide a very similar level of predictiveness.

The correct description of 2-point QCD correlators at large Euclidean momentum $Q$ (perturbative logarithm plus
power corrections in $Q^{-2}$) is provided by the Soft-Wall (SW) holographic model~\cite{son2} (an earlier variant
was suggested in Ref.~\cite{andreev}). This model
describes the Regge-like radial meson spectrum that in the case of spin-1 mesons reads $m_n^2=\mu^2(n+1)$, where $n=0,1,2,\dots$.
The SW model can be generalized towards inclusion of arbitrary intercept parameter $b$, $m_n^2=\mu^2(n+1+b)$~\cite{genSW}.
The necessity for non-zero intercept parameter $b$ is provided by the spectroscopy of light mesons. For example,
consider the non-strange isosinglet states in the vector and axial-vector channels.
According to the Particle Data~\cite{pdg}, there are only three well-established
$\omega$-mesons: $\omega(782)$, $\omega(1420)$, and $\omega(1650)$.
Taking their masses from~\cite{pdg} and ascribing them the "radial"
quantum numbers $n=0,1,2$, we obtain the fit (in GeV$^2$): $m^2_{\omega}(n)\approx1.1(n+0.7)$.
In the axial-vector sector, there is only one well-established
state $f_1(1285)$ (another one, $f_1(1420)$, consists mostly of the
strange quarks). As a motivated guess, let us use the non-confirmed
states $f_1(1970)$ and $f_1(2230)$~\cite{pdg}. We ascribe them
the "radial" quantum numbers $n=2,3$ (the state corresponding to
$n=1$ --- the isoscalar partner of $a_1(1640)$~\cite{pdg} --- is not
known). This gives the fit: $m^2_{f_1}(n)\approx1.1(n+1.5)$.
Thus we see that the correction to the vector intercept of standard SW model is at the level of 50\%.
Note a remarkable fact that the slopes in the obtained linear fits are approximately equal.

The holographic sum rules yield definite predictions for the intercept $b$ in the vector and axial case.
The aim of the present paper is to demonstrate how almost the same numerical predictions can be obtained in the opposite
limit of low $Q^2$ expansion of correlators. This looks really surprising since the corresponding equations
are very different from the case of high $Q^2$ expansion.
In addition, we will reproduce an expression for the slope $\mu^2$ known from QCD sum rules.

As a byproduct of our analysis, we will show how the effects of CSB can be embedded into holographic models
on the level of 2-point correlators, {\it i.e.} in agreement with (i). The usual bottom-up holographic description
of CSB is based on merging of chiral effective field theory with AdS/QCD models~\cite{son1,pom} that inevitably
includes higher n-point functions. Such a phenomenological approach has been remarkably successful
in the case of Hard Wall (HW) holographic model~\cite{son1,pom} (this model, however, does not reproduce
the Regge-like spectrum and power-like corrections in high $Q^2$ expansion of correlators) but faced problems
in simple realizations of SW ones. We will effectively
take into account some CSB effects\footnote{To be precise, we mean the mass splitting between the chiral partners, {\it e.g.},
between the vector and axial-vector mesons, which is induced by the CSB, and the difference in the low-energy behavior
of corresponding correlators. There is still not model-independent understanding of how these mass splittings are related
with other features of CSB, such as the Gell-Mann-Oakes-Renner relation and low energy theorems. The incorporation of these
features into our approach is a interesting task for the future.}
via different conditions on 2-point correlators at
zero momentum using some physical motivations. Within this scheme, even the simplest SW holographic
setup becomes a working model, at least after a certain reformulation.

The aforementioned reformulation
constitutes another one, albeit secondary, objective of our work. We will argue
that the replacement of enigmatic "dilaton background" by 5D mass depending on the fifth
coordinate looks more natural from a physical viewpoint and leads to a simpler model.
The proposals to introduce similar infrared modifications of 5D mass in AdS/QCD models have
appeared in the literature from time to time since Ref.~\cite{forkel}, we will motivate and
exploit this formulation in a more systematic manner.

The present paper is aimed at analysis of conceptual points mentioned above which may help in further
development of the theory of holographic QCD. We will not attempt to describe the meson spectrum
"in the best possible way" --- that problem is different in scope and often requires introduction
of some non-linearities to Regge trajectories (see, {\it e.g.}, the recent Refs.~\cite{nonl}; the non-linear radial
meson spectra within the SW holographic approach were analyzed in Ref.~\cite{nonlSW}).

The paper is organized as follows. In order to make our analysis and arguments self-contained,
in Section~2 we first briefly review the idea of AdS/QCD approach, reformulate the SW model and
reproduce predictions for Regge spectral parameters of vector mesons from high $Q^2$ expansion
of 2-point correlator. The main results are contained in Section~3, where near the same predictions
are obtained from the low $Q^2$ expansion. Some relevant discussions are given in Section~4
and we conclude in Section~5.

\section{Two-point vector correlators and OPE}

We first recall briefly the formalism of effective action and its holographic realization.
Within the functional approach to quantum field theory, the primary object is the partition
function $Z[\phi]$ of Green functions which has the physical meaning of vacuum-to-vacuum
transition amplitude in the presence of external source $\phi$,
$Z[\phi]=\langle 0_\text{out}|0_\text{in}\rangle_\phi$. The effective action $S_\text{eff}\{\phi\}$ is defined by
$Z[\phi]=\exp\left(iS_\text{eff}\{\phi\}\right)$. The coefficients of expansion of $S_\text{eff}\{\phi\}$
in the external field $\phi$ are the connected correlation functions of currents $J$ coupled to $\phi$.
In the momentum space, the effective Lagrangian density of Lorentz invariant theory is
\be
\mathcal{L}_\text{eff}=\frac12\hat{P}\text{Tr}\left[\phi\Pi_\phi(q^2)\phi\right]+\mathcal{O}\left(\phi^3\right),
\ee
where the term linear in $\phi$ disappears due to equation of motion, $\hat{P}$ denotes the corresponding
polarization tensor if the field has Lorentz indices, and the scalar function $\Pi_\phi$ is called
the two-point correlator of currents $J$ coupled to $\phi$ and is related to the full Green function of a
field theory via
\be
\langle JJ\rangle_\phi=\hat{P}\Pi_\phi.
\ee
In the large-$N_c$ limit of QCD, the correlator $\Pi_\phi$ is a meromorphic function and the higher n-point functions vanish~\cite{wit}.
It means in particular that in the limit $N_c\rightarrow\infty$,
$\Pi_\phi$ has the structure of sum over infinite number
of pole terms corresponding to contributions of infinitely narrow hadrons with quantum numbers of field $\phi$,
\be
\label{sum}
\Pi_\phi(q^2)\sim\sum_{n=0}^\infty\frac{F_n^2}{q^2-m_n^2},
\ee
where contact terms needed for regularization are omitted.

In the bottom-up holographic approach, following the ideas of AdS/CFT correspondence one assumes the existence
of 5D dual theory (in the sense of strong-weak duality) for QCD in the large-$N_c$ limit and tries to build a phenomenologically
useful gravitational 5D model. This putative 5D theory is constructed
in 5D Anti-de Sitter (AdS$_5$) space or asymptotically AdS$_5$ space. The well-known reason for the choice of this space
(at least when approaching boundary) is that only in this case the holographic principle for 4D field theories has chances to be naturally implemented ---
the 4D boundary of AdS$_5$ is time-like (and only in the case of this space if we wish to have a homogeneous space of constant curvature)
and has the appearance (at least after appropriate choice
of coordinates) of 4D Minkowski flat space. Also the AdS$_5$ space emerges naturally if one tries to "geometrize" the dilatation symmetry in a 4D conformal theory,
i.e., to rewrite the scaling transformations of field theory operators as space-time transformations of some 5D fields~\cite{sundrum}. For more specific motivations
which are relevant to QCD see Ref.~\cite{pol}. A convenient parametrization of the AdS$_5$ metric is given by the Poincar\'{e} patch with the line element
\be
\label{metric}
ds^2=\frac{R^2}{z^2}\left(\eta^{\mu\nu}dx_\mu dx_\nu-dz^2\right),
\ee
where $R$ is the radius of AdS$_5$ space and $z\geq0$ represents the fifth holographic coordinate that has
the physical meaning of inverse energy scale. The 4D Minkowski space becomes the ultraviolet boundary
of AdS$_5$ residing at $z=0$. The relation between a 4D gauge theory and its dual 5D gravitational theory
schematically is given by a concise statement
\be
\label{pres}
S_\text{eff}\{\phi(x)\}+J(x)\phi(x)=\left.S_\text{5D}^\text{boundary}\left(\phi(x,z)\right)\right|_{\phi(x,0)\doteq J(x)},
\ee
where $\phi(x,0)$ represents the UV boundary value of $\phi(x,z)$.
If the 4D gauge theory is in the strong coupling regime, the 5D theory must be weakly coupled due to
strong-weak duality. This general idea paved the way for building semiclassical 5D models
which describe the low energy QCD and are often interpolated to higher energies.

We will analyze the case of transverse vector fields, the corresponding polarization tensor is
\be
\hat{P}=\eta_{\mu\nu}-\frac{q_\mu q_\nu}{q^2}.
\ee
The simplest action of SW holographic model for vector mesons can be written as ($M,N=0,1,2,3,4$)
\be
\label{action}
S_\text{5D}=\frac{1}{g_5^2}\int d^4x\,dz\sqrt{g}\,e^\varphi\left(-\frac14 F^{MN}F_{MN}+\frac12m_5^2V^NV_N\right),
\ee
where $F_{MN}=\partial_MV_N-\partial_NV_M$ and $\varphi$ is given below in~\eqref{dilaton}.
The usual conditions,
\be
\label{restr}
\partial^\mu V_\mu=0, \qquad V_z=0,
\ee
are imposed on physical $4D$ excitations.
The 5D coupling $g_5$ plays the role of normalization constant for the field $V_M$.
The quadratic in field structure of action~\eqref{action} ensures disappearance of three and higher
point correlation functions as expected in the large-$N_c$ QCD.

In the bottom-up holographic constructions, the 5D fields are required to disappear at four-dimensional infinity $x_\mu\rightarrow0$
but extend to "holographic infinity" $z\rightarrow0$
(that corresponds to infinity along the holographic radial coordinate $r$ related to $z$
as $r=R^2/z$), i.e. the fields are to live not only in the 5D bulk but also on the UV boundary $z=\epsilon$,
$\epsilon\rightarrow 0$. They are also required to vanish on the IR boundary $z=\infty$. With these requirements imposed, one can integrate~\eqref{action}
by parts and arrive at the action
\begin{multline}
\label{action2b}
S_\text{5D}=\frac{1}{2g_5^2}\int d^4x\,dz\, V_M\biggl\{\partial_P\Bigl[\sqrt{g}\,e^\varphi\left(g^{PS}g^{MN}-g^{PM}g^{SN}\right)\partial_S\Bigr]\\
+m_5^2\sqrt{g}\,e^\varphi g^{MN}\biggr\}V_N
-\frac{1}{2g_5^2}\left.\int d^4x\sqrt{g}\,e^\varphi g^{zz}g^{MN} V_M\partial_zV_N\right|_{z=\epsilon}.
\end{multline}
The dynamics in the dual 5D theory are governed by the classical Equation Of Motion (e.o.m.) that
leaves only the surface term in the action~\eqref{action2b},
\be
\label{action2c}
S_\text{on-shell}=-\frac{1}{2g_5^2}\left.\int d^4x\sqrt{g}\,e^\varphi g^{zz}g^{MN} V_M\partial_zV_N\right|_{z=\epsilon}.
\ee

The AdS/CFT prescription for
5D mass of $p$-form field reads~\cite{witten}
\be
m_5^2R^2=(\Delta-p)(\Delta+p-4),
\ee
where $\Delta$ is the
canonical dimension of operator dual to the field $V_M$ on the AdS$_5$ boundary.
For vector fields the prescription dictates
\be
\label{mass}
m_5^2R^2=(\Delta-1)(\Delta-3),
\ee
since the vector field is a $p=1$ form.
The usual quark vector current $\bar{q}\gamma_\mu q$ has $\Delta=3$, hence, $m_5^2=0$. Vector operators of higher dimensions
correspond to massive 5D vector fields.

The dilaton-like quantity $\varphi$ in the action~\eqref{action} dictates a background.
The standard SW holographic model is defined by~\cite{son2}
\be
\label{dilaton}
\varphi=cz^2,
\ee
where the constant $c$ provides a mass scale and can be both positive and negative.
The choice~\eqref{dilaton} yields the Regge-like spectrum~\cite{son2},
\be
\label{SWspectrum}
m_n^2=4|c|(n+1), \qquad n=0,1,2,\dots,
\ee
and what is very important it leads to a correct analytical structure of OPE.
The physical origin of~\eqref{dilaton} remains an open problem. As was first noticed
in Ref.~\cite{No-wall}, the dilaton-like background can be absorbed into a infrared
modification of $m_5^2$. Indeed, using the metric~\eqref{metric} and restriction~\eqref{restr}
the non-vanishing term with $z$-derivative in action~\eqref{action} can be written as
\be
\sqrt{g}\,e^{cz^2}\partial^zV^N\partial_zV_N=\left(\frac{R}{z}\right)^5e^{cz^2}g^{zR}g^{NS}\partial_RV_S\partial_zV_N
=\frac{R}{z}e^{cz^2}\left(\partial_zV_N\right)^2,
\ee
where the inverse to metric $g_{MN}$ in~\eqref{metric} tensor is $g^{MN}=\frac{z^2}{R^2}\eta^{MN}$
and we write only lower indices when contraction with flat metric is understood, {\it e.g.}, $V_N^2=\eta^{MN}V_MV_N$.
Now we redefine the field
\be
V_N=e^{-cz^2/2}v_N,
\ee
and arrive at
\be
\frac{R}{z}\left(\partial_zv_N-czv_N\right)^2=
\sqrt{g}\left(\partial^zv^N\partial_zv_N+\frac{c^2z^4}{R^2}v^Nv_N\right)-2Rcv_N\partial_zv_N.
\ee
We see that a $\mathcal{O}(z^4)$ mass term emerged. The last term will not contribute
to the e.o.m.. It is easy to show, however, that for scalar and tensor cases the analogous term
contains a $z$-dependent factor and does contribute to the e.o.m. resulting in $\mathcal{O}(z^2)$
mass term.

The standard SW model can be thus reformulated without $z$-dependent dilaton
background~\eqref{dilaton} if we instead use the following ansatz for $z$-dependent
5D mass,
\be
\label{ansatz}
m_5^2(z)R^2=a+bz^2+c^2z^4,
\ee
which stays in the action
\be
\label{action2}
S_\text{5D}=\frac{1}{2g_5^2}\int d^4x\,dz\sqrt{g}\Bigl(-\partial^Mv^N\partial_Mv_N+m_5^2(z)v^Nv_N\Bigr),
\ee
Any infrared modifications of dilaton background or metric can be translated
into corresponding modifications of~\eqref{ansatz}. The constant $a$ must be
identified with the AdS/CFT prescription~\eqref{mass}, the further terms are infrared corrections.
It should be emphasized that the given corrections do not violate this prescription
because it is formulated on the UV boundary, $z\rightarrow 0$, only (the Ref.~\cite{sundrum} contains
a nice discussion of this point). For this reason any
correction to~\eqref{mass} disappearing in the limit $z\rightarrow 0$ is compatible
with the AdS/CFT prescription. On the other hand, arbitrary modifications of $z$-dependence in~\eqref{ansatz}
will, generally speaking, spoil the structure of standard OPE for two-point correlators.
This issue is closely related with the
known fact that non-linear corrections to Regge-like spectrum~\eqref{SWspectrum}
after summation over resonances in~\eqref{sum} generically lead to analytical structures incompatible
with the OPE~\cite{we}. The compatibility can be achieved only if
such corrections decrease with $n$ exponentially or faster~\cite{we,npb}. A form of potential
in the corresponding e.o.m. (written in a Schr\"{o}dinger-like form)
that would generate these corrections is unknown. In the present study, we will adhere
to the ansatz~\eqref{ansatz} describing the linear spectrum in the most general form.
Aside from reproducing correctly the analytical structure of OPE, the case of linear spectrum
has the advantage of being exactly solvable.

For phenomenological description of real meson spectra with different quantum numbers
one should introduce different intercepts in the spectrum~\eqref{SWspectrum}.
Within the SW model, this can be achieved by different infrared modifications of dilaton
background or AdS metric. This fine tuning looks as if we constructed different dual models
for different quantum numbers. The reformulation above looks nicer in this respect:
A dual holographic model is unique but infrared modifications of $m_5^2$ are different.
Note that the parameter $c$ in~\eqref{ansatz} is independent of quantum numbers since
it dictates the slope of radial trajectories which phenomenologically is indeed
approximately universal~\cite{phen,phen2,phen3,phen4,phen5,phen6}. The given universality seems to be
an important consequence of confinement and appears naturally
in hadron string approaches and some related quark models~\cite{linear,linear2,linear3,linear4,linear5,linear6,linear7,linear8,linear9,linear10}.
The intercept will be determined by parameters $a$ and $b$.
The first one can be fixed by~\eqref{mass}. One of our goals will be determination of the
second intercept parameter $b$.

The e.o.m. ensuing from the action~\eqref{action2} with $m_5^2(z)$
from~\eqref{ansatz} after 4D Fourier transform
\be
\label{ft}
v_\mu(q,z)=\int d^4x\,e^{iqx}v_\mu(x,z),
\ee
takes the form
\be
\label{eom}
\left[-q^2-z\partial_z\left(\frac{1}{z}\partial_z\right)+\frac{m_5^2(z)R^2}{z^2}\right]v_\mu(q,z)=0.
\ee
In the present work, we will consider the case $a=0$ in~\eqref{ansatz}. This is an
important case of twist 2 vector current according to relation~\eqref{mass} and is the most
studied in the literature\footnote{The twist 2 operators, $\tau=\Delta-J=2$, correspond to operators
of lowest scaling dimension at a fixed spin $J$. Usually they play the decisive role in QCD dynamics.
For instance, they give the dominant contribution in the analysis of hadronic deep inelastic scattering via the OPE~\cite{br}.
Also they interpolate the hadron states in the QCD sum rules~\cite{svz} and in the lattice simulations. In the conformal field
theory, they correspond to primary operators, i.e., realize the irreducible representations of conformal group and, thus, according to the
AdS/CFT correspondence, are dual to 5D fields falling into irreducible representations of AdS$_5$ space-time symmetry.}.
In order to make comparisons with the OPE in QCD, we must calculate
the two-point correlator in Euclidean space, i.e. introducing Euclidean momentum $Q^2=-q^2$.
Following the standard holographic procedure, we should find the solution of Eq.~\eqref{eom}
in the form
\be
\label{form}
v_\mu(q,z)=v_\mu(q)v(q,z),
\ee
with the boundary condition $v(q,0)=1$. Then
$v_\mu(q)$ can be interpreted as the source. The corresponding
solution for the scalar shape function $v(q,z)$, satisfying also $v(q,\infty)=0$,
is (we pass to Euclidean space in what follows)
\be
\label{sol}
v(\mathcal{Q},\zeta)=\Gamma\left(1+\mathcal{Q}^2+\beta\right)e^{-\zeta^2/2}U\left(\mathcal{Q}^2+\beta,0;\zeta^2\right),
\ee
where $U$ is the Tricomi confluent hypergeometric function and for simplicity of further relations
we defined the dimensionless quantities
\be
\mathcal{Q}^2=\frac{Q^2}{4|c|},\qquad \beta=\frac{b}{4|c|},\qquad \zeta^2=|c|z^2.
\ee
Note that the Eq.~\eqref{eom} has the second solution which, however, diverges as $\zeta\rightarrow\infty$,
\be
v_2(\mathcal{Q},\zeta)\sim \zeta^2e^{-\zeta^2/2}L^1_{-(1+\mathcal{Q}^2+\beta)}(\zeta^2)\sim e^{\zeta^2/2}\zeta^{2(\mathcal{Q}^2+\beta)}.
\ee
Here $L^1_{S}(x)=(1+S)\,{}_1F_1\left(-S,2,x\right)$ denotes the corresponding Laguerre function.
We discard this solution since
it does not satisfy the holographic "regularity in the bulk" condition~\cite{witten}.
The solution~\eqref{sol} is regular in the bulk as
$U\left(\mathcal{Q}^2+\beta,0;\zeta^2\right)\sim \zeta^{-2(\mathcal{Q}^2+\beta)}$ at large $\zeta$.

Inserting~\eqref{form} into the action~\eqref{action2c} and taking the
second functional derivative with respect to $v_\mu(q)$ we get the usual AdS/QCD expression
for the vector 2-point correlator in Euclidean space,
\be
\label{Vcor}
\Pi_V(\mathcal{Q}^2)=\lim_{\zeta\rightarrow 0}\left(-\frac{R}{g_5^2}\frac{\partial_\zeta v(\mathcal{Q},\zeta)}{\zeta}\right).
\ee
The expansion of solution~\eqref{sol} at $\zeta\rightarrow 0$ yields
\be
\label{expan}
v(\mathcal{Q},\zeta)=1+
\left\{\left(\mathcal{Q}^2+\beta\right)\Bigl[\ln\zeta^2
+\psi\left(1+\mathcal{Q}^2+\beta\right)+2\gamma-1\Bigr]-\frac12\right\}\zeta^2,
\ee
where $\psi$ denotes the digamma function and $\gamma\approx0.577$ is the Euler's constant.

Following the standard procedure, we should substitute~\eqref{expan} into~\eqref{Vcor}
and take the limit $\zeta\rightarrow 0$. The logarithmic term will give the divergent
contribution proportional to $\left(\mathcal{Q}^2+\beta\right)\ln\zeta^2$ which
is interpreted as a contact term and discarded. This regularization is defined up to
adding an arbitrary constant (or polynomial of $\mathcal{Q}^2$) which fixes the renormalization
scheme. The high-energy predictions are scheme-independent. We will be interested, however, in
low-energy predictions. In the latter case, adding arbitrary constant can change predictions
in an arbitrary manner. We should thus fix a specific renormalization scheme that is the most
adequate to the physical problem under consideration. Below we motivate such a scheme.

First of all, let us proceed more accurately following the relevant prescriptions of
Holographic Renormalization (HR)~\cite{skenderis}. The first step is to define the subtracted action as the
sum of regularized and counterterm actions,
\be
S_\text{sub}\left[v(q,\epsilon);\epsilon\right]=S_\text{reg}\left[v(q,0);\epsilon\right]
+S_\text{ct.}\left[v(q,\epsilon);\epsilon\right].
\ee
The counterterm action $S_\text{ct.}$ must absorb the arising infinities such that the renormalized on-shell
action $S_\text{ren}$ defined as
\be
S_\text{ren}\left[v(q,0)\right] = \lim_{\epsilon\rightarrow0} S_\text{sub}\left[v(q,\epsilon);\epsilon\right],
\ee
is finite. In order to obtain correlation functions
one varies $S_\text{sub}$, taking $\epsilon\rightarrow 0$ only at the end of the calculation. For this reason
the distinction between $S_\text{sub}$ and $S_\text{ren}$ is important.

With our definitions and notations, the action~\eqref{action2c} takes the form
\be
\label{action2d}
S=-\frac{R}{2g_5^2}\int d^4x\, v_\mu(\mathcal{Q}) v_\nu(\mathcal{Q})\eta^{\mu\nu}
\left(\frac{1}{\zeta}v(\mathcal{Q},\zeta)\partial_\zeta v(\mathcal{Q},\zeta)\right)_{\zeta=\epsilon}.
\ee
For finding the structure of $S_\text{ct.}$ we should rewrite the near-boundary expansion~\eqref{expan} as
\be
\label{expan2}
v(\mathcal{Q},\zeta)=v_{(0)}+\zeta^2\left(v_{(2)}+\tilde{v}_{(2)}\ln\zeta^2\right) + \mathcal{O}(\zeta^4).
\ee
According to the HR formalism~\cite{skenderis}, the coefficients $v_{(i)}$ (which in coordinate space represent certain
functions of $x$, $v_{(i)}=v_{(i)}(x)$) have the following interpretation. $v_{(0)}$ is the field theory source.
$\tilde{v}_{(2)}$ is related to conformal anomaly; it is a local function of $v_{(0)}$ and can be determined by
the near-boundary analysis of e.o.m. The coefficient $v_{(2)}$ is a non-local function of $v_{(0)}$ in the coordinate space
(they cannot be related by taking a finite number of derivatives),
it is undetermined by the near-boundary analysis of e.o.m. and can be obtained only from the full solution in the bulk.

The next step is to substitute~\eqref{expan2} back to~\eqref{action2d}, the key factor will be
\be
\label{key}
\left(\frac{1}{\zeta}v\partial_\zeta v\right)_{\zeta=\epsilon}=2v_{(0)}\left(v_{(2)}+\tilde{v}_{(2)}+\tilde{v}_{(2)}\ln\epsilon^2\right)+\dots
\ee
This expression dictates the structure of the counterterm action --- it must cancel the logarithmic divergence
near the UV boundary. Thus we obtain
\be
\label{ct}
S_\text{ct.}=\frac{R}{g_5^2}\int d^4x\, v_\mu(\mathcal{Q}) v_\nu(\mathcal{Q})\eta^{\mu\nu} v_{(0)}\tilde{v}_{(2)}\ln\epsilon^2 + \text{Const},
\ee
where a choice of Const fixes the renormalization scheme. Strictly speaking, one should invert the relations
$v_{(0)}(v)$ and $\tilde{v}_{(2)}(v)$ and write $S_\text{ct.}$ in terms of $v$ because in the coordinate space it is
$v(x,\epsilon)$ that defines the field $v_\mu$ (see~\eqref{ft} and~\eqref{form}) transforming as a vector under bulk
diffeomorphisms at $\zeta=\epsilon$. But this detail will not be essential for us.

In the minimal subtraction scheme, the subtracted action becomes thus proportional to
$v_{(0)}\left(v_{(2)}+\tilde{v}_{(2)}\right)$ according to~\eqref{key}. There is still the freedom to add a finite counterterm.
This corresponds to scheme dependence in the field theory. It may happen that a physical problem under consideration
dictates unambiguously which finite counterterm should be added~\cite{skenderis}. We are of the opinion that this is relevant
for our model as well. Let us recall a physical interpretation that is often ascribed to the subtraction of UV infinities
in the field theory: One subtracts the contribution of a region lying outside the region of validity of the field theory,
{\it i.e.} where the field theory needs a UV completion. A similar philosophy can be applied to the bottom-up holographic model ---
in the ultraviolet limit $z\rightarrow0$, the applicability of semiclassical holographic models to QCD looks questionable and
the whole approach seems to need modifications (see, {\it e.g.}, discussions in Refs.~\cite{UV,UV2,UV3}). Keeping this in mind, we observe that
the contribution from finite term $\tilde{v}_{(2)}$ also emerges from that UV region, hence, it is reasonable to subtract
the given finite term as well\footnote{We would remark also that subtraction of both terms proportional to $\tilde{v}_{(2)}$
removes a potential double counting in the following sense: These terms are related with conformal anomalies~\cite{skenderis}, {\it i.e.}
with generation of a scale at a quantum level, but a scale is already introduced by hand in the model.}.
This fixes our subtraction scheme: We retain only term $v_{(2)}$ that arises from dynamics in the
bulk and subtract all terms stemming from the UV region where our model is viewed us inapplicable. All constants appearing in $v_{(2)}$
are considered as prediction of the model --- we cannot modify them manually since this would cause an arbitrary change of
low-energy predictions.

On the operational level, the UV diverging contribution proportional to $\zeta^2\ln\zeta^2$
appears because the Tricomi function $U(S,0;z^2)$ is not holomorphic in the point $z=0$.
Our regularization is equivalent to subtraction of this singularity, the regularized solution becomes holomorphic at $z=0$.

Inserting the regularized expansion~\eqref{expan} into~\eqref{Vcor} we obtain the two-point vector
correlator in our model,
\be
\label{Vcor2}
\Pi(\mathcal{Q}^2)=-\frac{2R}{g_5^2}|c|\left\{\left(\mathcal{Q}^2+\beta\right)
\left[\psi\left(1+\mathcal{Q}^2+\beta\right)+2\gamma-1\right]-\frac12\right\}.
\ee
This correlator is plotted in Fig.~1 for some typical values of parameters.
\begin{figure}[!ht]
  \center{\includegraphics[width=0.9\linewidth]{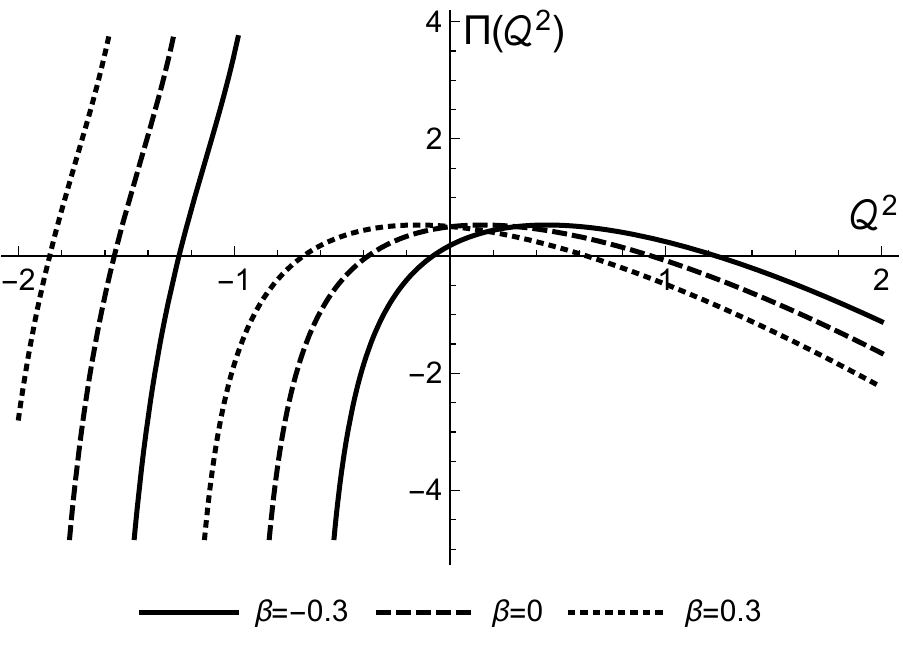}}
  \caption{\footnotesize The two-point vector
correlator~\eqref{Vcor2} for three typical values of intercept parameter $\beta$ (see the discussions after Eq.~\eqref{slope}). The general factor $\frac{2R|c|}{g_5^2}$ is set to unity.
In the time-like region of dimensionless momentum squared $\mathcal{Q}^2$, the plot extends to the position of the first pole (see~\eqref{sp}).}
\end{figure}\\
The expression~\eqref{Vcor2} at $\beta=0$ coincides with the result of
"No-wall" holographic model of Ref.~\cite{No-wall}. In fact this expression can be obtained
directly from the $\beta=0$ case by observing that $b\neq0$ corresponds just
to the shift of momentum squared $q^2\rightarrow q^2-b$ in the e.o.m.~\eqref{eom},
{\it i.e.} $\mathcal{Q}^2\rightarrow\mathcal{Q}^2+\beta$.

The digamma function in~\eqref{Vcor2} has poles at
\be
\label{sp}
-\mathcal{Q}^2_n=\frac{m_n^2}{4|c|}=n+1+\beta,\qquad n=0,1,2,\dots,
\ee
which correspond to the mass spectrum of the model.
The substitution of known pole representation for this function,
\be
\psi(1+x)=\sum_{n=1}^{\infty}\frac{1}{n}-\sum_{n=1}^\infty\frac{1}{n+x}-\gamma,
\ee
into~\eqref{Vcor2} leads to the expected analytical structure~\eqref{sum} written
in the Euclidean space.

With the help of asymptotic representation of digamma function for large argument $x\rightarrow\infty$,
\be
\psi(1+x)\simeq\ln x+\frac{1}{2x}-\frac{1}{12x^2}+\mathcal{O}(x^{-3}),
\ee
we can expand~\eqref{Vcor2} at large $\mathcal{Q}^2$,
\be
\label{Vcor3}
\Pi(\mathcal{Q}^2)\simeq-\frac{2R}{g_5^2}|c|\left\{\left(\mathcal{Q}^2+\beta\right)\ln\mathcal{Q}^2
+\frac{\beta^2-1/6}{2\mathcal{Q}^2}+\frac{\beta(1-2\beta^2)}{12\mathcal{Q}^4}+\dots\right\},
\ee
where the contact and $\mathcal{O}(\mathcal{Q}^{-6})$ terms are omitted.
The expansion~\eqref{Vcor3} only partly coincides with the result of generalized SW model of Ref.~\cite{genSW},
where the arbitrary intercept $\beta$ in the linear spectrum~\eqref{sp} was introduced
via the modified dilaton background $e^{cz^2}\rightarrow U^2(\beta,0;cz^2)e^{cz^2}$.

The relevant form of OPE in QCD in the chiral and large-$N_c$ limits reads~\cite{svz}
\be
\label{OPE}
\Pi^{OPE}(\mathcal{Q}^2)\simeq-\frac{N_c}{6\pi^2}|c|\mathcal{Q}^2\ln\mathcal{Q}^2+
\frac{\langle\frac{\alpha_s}{\pi} G^2\rangle}{96|c|\mathcal{Q}^2}+
\frac{\xi}{9}\frac{\pi\alpha_s\langle\bar{q}q\rangle^2}{8c^2\mathcal{Q}^4}+\dots,
\ee
where $\langle\frac{\alpha_s}{\pi} G^2\rangle$ and $\langle\bar{q}q\rangle$ are
the gluon and quark condensates, respectively, and $\xi$ depends on the space and charge parities
($\xi=-7$ for the vector case and $\xi=11$ for the axial-vector one).
The omitted constant contribution in~\eqref{OPE} (part of contact terms) allows
to change the renormalization scale $\mu$ in the perturbative logarithm,
we used this freedom to set $\mu^2=4|c|$.

The expansions~\eqref{Vcor3} and~\eqref{OPE} can be now matched.
The matching of coefficients in front of the leading logarithms
gives the standard normalization factor for the 5D vector fields,
\be
\label{norm}
\frac{R}{g_5^2}=\frac{N_c}{12\pi^2}.
\ee
The matching of $\mathcal{O}(\mathcal{Q}^{-2})$ terms relates the gluon condensate
to the spectral parameters as
\be
\label{gl}
\left\langle\frac{\alpha_s}{\pi} G^2\right\rangle=\frac{N_c}{2\pi^2}\left(\frac16-\beta^2\right)(4c)^2.
\ee
As was shown in Ref.~\cite{genSW}, a good agreement for the phenomenological slope of
radial trajectories~\cite{phen,phen2,phen3,phen4,phen5,phen6}
\be
\label{slope}
4|c|\approx1.2\,\text{GeV}^2
\ee
is achieved for the intercept parameter $|\beta|\approx0.3$, where $\beta=\mp0.3$ refers to
the vector and axial cases, correspondingly. The arguments were the following: (i) $|\beta|\approx0.3$
leads to a correct value of the gluon condensate in relation~\eqref{gl},
$\langle\frac{\alpha_s}{\pi} G^2\rangle\approx(360\,\text{MeV})^4$;
(ii) $\beta\approx-0.3$ gives a satisfactory description the spectrum of radially excited $\omega$-mesons
within the accuracy of large-$N_c$ limit;
(iii) $\beta\approx-0.3$ reproduces the observed electromagnetic decay width of $\omega$-meson;
(iv) $\beta\approx0.3$ describes reasonably the spectrum of radially excited axial $f_1$-mesons.
The description of spectrum of excited isovector $\rho$ and $a_1$ mesons with $|\beta|\approx0.3$ is even better.
In addition, a remarkable qualitative agreement takes place: The first non-perturbative correction
to the parton logarithm does not depend on parity while the next one is of opposite sign for different
parities.

The quark effects are not incorporated into the holographic model under consideration. For this reason
it is not surprising that the quark contributions like $\mathcal{O}(\mathcal{Q}^{-4})$ and
$\beta\ln\mathcal{Q}^2$ do not match. The former can have a sizeable (or even dominating
in the large-$N_c$ limit) contribution from dimension 6 gluon condensate $\varepsilon^{abc}\langle G_aG_bG_c\rangle$
which is not present in the phenomenological expansion~\eqref{OPE}. The latter appears after taking
into account non-zero bare quark masses in the quark loops~\cite{svz}, i.e. from explicit breaking
of chiral symmetry. Our model thus predicts that at low momenta the parton model logarithm in~\eqref{OPE}
acquires a contribution related with spontaneous chiral symmetry breaking (that might be related with formation of
effective constituent quark mass) which is absent in the standard
OPE constructed starting from the perturbation theory.

\section{Predictions from correlators at zero momentum}

Now we will demonstrate how the same prediction $|\beta|\approx0.3$ can follow
from a simple analysis of vector correlator at zero momentum. Our key proposal is to treat
the linear spectrum of SW model~\eqref{SWspectrum} as dual (in the sense of quark-hadron duality) to
perturbation theory corrected by gluonic power terms in the OPE~\eqref{OPE}. These terms
are known to appear from breaking of conformal symmetry. In other words,
the spectrum~\eqref{SWspectrum} does not yet correspond to real resonances but appears
as a result of "meromorphization" of perturbative background corrected at low energies by
conformal symmetry breaking power terms. The hadron resonances are associated with deviations
from a background. In our model, this means then that they are associated with non-zero intercept
parameter\footnote{In other models, the spectrum corresponding to perturbative background can be
different. For instance, the spectrum of $S$-wave spin-1 mesons in light front holographic QCD
behaves as $m_n^2\sim n+1/2$~\cite{br}, {\it i.e.} one should look for deviations from $\beta=-1/2$.
It is curious to observe that this spectrum minimizes the power contributions to perturbative
logarithm in the 2-point vector correlator within the class of linear spectra~\cite{npb}.
The value of $\beta=-1/2$, however, is incompatible with positivity of gluon condensate~\eqref{gl}
in our approach.} $\beta$ in~\eqref{sp}.

Simultaneously $\beta\neq0$ signifies the breaking of chiral symmetry ---
the appearance of contributions from quark condensate in the expansion~\eqref{Vcor3} and of mass splitting between
parity partners.

The given philosophy can be converted into a predictive calculational scheme. Consider
the vector correlator~\eqref{Vcor2} at zero momentum,
\be
\label{Vzero}
\Pi(0)=-\frac{2R}{g_5^2}|c|\left\{-\frac12+\beta\Bigl[\psi\left(1+\beta\right)+2\gamma-1\Bigr]\right\}.
\ee
It is important to emphasize the role of our regularization: $\Pi(0)$ is a finite quantity with all constants fixed.
In the standard regularization, the subtraction of infinite constant makes $\Pi(0)$ ambiguous.

The successful old hypothesis of Partial Conservation of Axial Current predicts in the chiral limit
that the value of axial-vector correlator is
\be
\label{Azero}
\Pi_A(0)=f_\pi^2,
\ee
where $f_\pi$ is the weak pion decay constant emerging from the pion pole. In essence, one may
regard~\eqref{Azero} as an alternative definition of $f_\pi$ --- if the spontaneous chiral symmetry
breaking is assumed, the r.h.s. of~\eqref{Azero} must give the residue of the corresponding Goldstone
boson pole. The critical observation is that
the axial-vector resonances should not contribute to $\Pi_A(0)$ because of a large mass gap
in the axial channel --- $f_\pi^2$ absorbs effectively all contributions to $\Pi_A(0)$.
According to our philosophy, this means that terms with $\beta$ do not contribute.
We get thus from~\eqref{Vzero} the equation for the axial intercept parameter $\beta_a$,
\be
\label{ax}
\beta_a\Bigl[\psi\left(1+\beta_a\right)+2\gamma-1\Bigr]=0.
\ee
This equation has two numerical solutions: $\beta_a=0$ (no CSB as $\langle\bar{q}q\rangle=0$, see the last term in~\eqref{Vcor3})
and $\beta_a\approx0.31$ which is almost exactly the value extracted above from the large $Q^2$ limit.

When the equation~\eqref{ax} holds, we can obtain from~\eqref{Vzero},~\eqref{Azero} and~\eqref{norm}
the following relation for the slope of linear spectrum~\eqref{sp},
\be
\label{slope2}
4|c|=\frac{48\pi^2}{N_c}f_\pi^2.
\ee
This relation for linear spectrum was derived in QCD sum rules in the large-$N_c$ limit
under various assumptions\footnote{The relation~\eqref{slope2} can be simply obtained~\cite{npb} by
combining the slope $2m_\rho^2$ of spectra of Veneziano-like dual amplitudes with the relation
$m_\rho^2=(24\pi^2/N_c)f_\pi^2$ which often holds in models respecting the Vector Meson Dominance~\cite{vmd}
including the classical QCD sum rules~\cite{svz}. It also emerges in attempts to incorporate the spontaneous CSB
into the hadron string framework~\cite{Andrianov}.} (see, {\it e.g.}, Refs.~\cite{lin}).
The relation~\eqref{slope2} meets well the phenomenology ---
in the real world with $N_c=3$, the empirical value~\eqref{slope} is reproduced for $f_\pi=87$~MeV,
which is the value of $f_\pi$ in the chiral limit according to the chiral perturbation theory~\cite{gasser}.
We get the relation~\eqref{slope2} in the opposite to the OPE based sum rules limit $Q^2\rightarrow0$.
In principle, we could act in the reverse direction --- require~\eqref{slope2} and obtain Eq.~\eqref{ax}
as a consequence.

In the vector channel, the value of $\Pi_V(0)$ is also non-zero but the physical reason must be
completely different --- a conversion of $\omega$ and neutral $\rho$ mesons into massless photons and
back is possible (the effect underlying the famous hypothesis of Vector Meson Dominance) leading to a kind of
effective "photon" contribution. We do not know this
contribution {\it apriori}, however, we can use the relation~\cite{ecker}
(a form of the "Das-Mathur-Okubo sum rule"~\cite{DMO})
\be
\label{L10}
-4L_{10}=\frac{d}{dQ^2}\left.\left(\Pi_V-\Pi_A\right)\right|_{Q^2=0},
\ee
where $L_{10}$ is one of constants of $SU_f(3)$ chiral Lagrangian~\cite{gasser,Pich}. From~\eqref{Vcor2}
we get
\be
\label{V0}
\frac{d}{dQ^2}\left.\Pi\right|_{Q^2=0}=-\frac{R}{2g_5^2}
\Bigl[\psi\left(1+\beta\right)+2\gamma-1+\beta\psi\left(1,1+\beta\right)\Bigr].
\ee
Using Eq.~\eqref{ax} and normalization~\eqref{norm} we finally obtain an equation for
the vector intercept parameter $\beta_v$,
\be
\label{V1}
\psi\left(1+\beta_v\right)+2\gamma-1+\beta_v\psi\left(1,1+\beta_v\right)-\beta_a\psi\left(1,1+\beta_a\right)
=\frac{96\pi^2L_{10}}{N_c}.
\ee
The typical values of $L_{10}$ extracted in the phenomenology lie near $L_{10}\approx-5.5\cdot10^{-3}$~\cite{Pich}. One should
keep in mind, however, that the chiral constants are scale dependent and the aforementioned values refer to the
scale of $\rho$-meson mass, $L_{10}(m_\rho)$. But we should substitute to Eq.~\eqref{V1} the value at zero momentum.
Fortunately the value of $L_{10}(0)$ can be determined in a scale-independent manner from hadronic $\tau$-decays.
The extracted value is $L_{10}(0)=(-6.36\pm0.09|_\text{expt}\pm0.16|_\text{theor})\cdot10^{-3}$~\cite{53}.
With this value\footnote{An order by magnitude estimation of $SU_f(3)$ chiral constants is
$L_i\sim\frac{f_\pi^2}{\Lambda^2_\text{CSB}}$~\cite{Pich}, where $\Lambda_\text{CSB}$ is the scale of spontaneous chiral
symmetry breaking. This scale is usually estimated from variation of contributions of chiral loops, the
result is $\Lambda_\text{CSB}\simeq4\pi f_\pi$~\cite{CSBscale} and leads to $L_i\sim1/(4\pi)^2\approx6.3\cdot10^{-3}$.
Note that the slope of radial trajectories~\eqref{slope2} at $N_c=3$ is just $\Lambda^2_\text{CSB}$.
The given coincidence certainly should have deep physical roots.} and $\beta_a=0.31$ the numerical solution
of Eq.~\eqref{V1} yields  $\beta_v\approx-0.26$ which is close to $\beta_v\approx-0.3$ estimated
above (the exact agreement $\beta_v=-\beta_a$ is achieved at $L_{10}=-7.5\cdot10^{-3}$).

An alternative strategy for making fits can consist in writing equation for $N_c=3$ from~\eqref{L10}
and~\eqref{V0},
\be
\psi\left(1+\beta_v\right)+\beta_v\psi\left(1,1+\beta_v\right)
-\psi\left(1+\beta_a\right)-\beta_a\psi\left(1,1+\beta_a\right)
=32\pi^2L_{10},
\ee
and imposing $\beta_a=-\beta_v$ to have a universal gluon condensate in the OPE~\eqref{Vcor3}.
This would give $\beta_a=-\beta_v\approx0.27$. After that we could reproduce a
reasonable value of gluon condensate and observe a very small contribution of terms with $\beta_a$
to the axial correlator at zero momentum (we omit the general factor in what follows),
\be
\Pi_A(0)\sim-\frac12+\beta_a\left[\psi\left(1+\beta_a\right)+2\gamma-1\right]\approx-\frac12-0.01,
\ee
while the corresponding contribution to the vector correlator would be relatively large,
\be
\Pi_V(0)\sim-\frac12+\beta_v\Bigl[\psi\left(1+\beta_v\right)+2\gamma-1\Bigr]\approx-\frac12+0.27.
\ee

The given simple calculations demonstrate phenomenologically how the constant contributions to
correlators which are interpreted as part of "contact" terms and neglected in high $Q^2$
expansions play a decisive role at low $Q^2$.

\section{Discussions}

There is a widespread opinion (see, {\it e.g.}, discussions in Ref.~\cite{holog2010}) that the phenomenology of bottom-up
holographic models has much
in common with QCD sum rules in the large-$N_c$ limit (sometimes called planar QCD sum rules) which were a fruitful
phenomenological tool in the past~\cite{sr,sr2,sr3,sr4,sr5,sr6,sr7,sr8,sr9,we,npb,lin}. Within this method, one takes the pole representation of two-point
correlators~\eqref{sum} in the Euclidean space,
\be
\Pi(Q^2)\sim Q^2\sum_{n=0}^\infty\frac{F_n^2}{Q^2+m_n^2},
\ee
assumes some ansatz for the spectrum, sums over all states, matches the result with the corresponding OPE
and low-energy constraints and finally derives phenomenological predictions. In the case of simple linear spectrum,
\be
m_n^2=\mu^2(n+b),
\ee
with constant residues $F_n^2$ one obtains the usual representation via digamma function,
\be
\label{st1}
\Pi(Q^2)\sim -\mathcal{Q}^2\psi(\mathcal{Q}^2+b)+\text{const},
\ee
where $\mathcal{Q}^2=Q^2/\mu^2$.

We wish to emphasize an important technical distinction between this approach and our holographic one.
In the vector case, we obtain a structure of the kind
\be
\label{st2}
\Pi(\mathcal{Q}^2)\sim -\left(\mathcal{Q}^2+\beta\right)\psi(\mathcal{Q}^2+1+\beta)+\text{const}.
\ee
Now changing intercept we change also the coefficient in front of $\psi$-function.
As a result, the expansion in large $\mathcal{Q}^2$ of~\eqref{st2} becomes different from the expansion
of~\eqref{st1} and consequently leads to a different set of sum rules when one matches to the OPE.
The reason of arising distinction lies in the fact that the holographic models represent a
dynamical approach where $\mathcal{Q}^2=-q^2/\mu^2$ appears in holographic e.o.m. like Eq.~\eqref{eom}:
Any shift of constant intercept $b\rightarrow b+\Delta b$ can be interpreted as
a shift $Q^2\rightarrow Q^2+\mu^2\Delta b$ in the corresponding e.o.m., {\it i.e.}
as a redefinition of what we call\footnote{Actually this remark is valid for any extension of Klein--Gordon
equation preserving the 4D relativistic invariance.} "$Q^2$". Physically this means, of course, a fine tuning
of mass gap. It is interesting to note that this shift ({\it i.e.} $\beta\neq0$ in~\eqref{st2})
leads to the appearance of contribution $\beta\ln\mathcal{Q}^2$ in the large $\mathcal{Q}^2$ expansion
(as in expansion~\eqref{Vcor3}). As we remarked in the end of Sect.~2, this contribution should be related with
the CSB in the form of acquired constituent quark mass. It is interesting to note that similar contributions were
interpreted as contributions stemming from the effective non-local dimension-two gluon condensate in Refs.~\cite{dim2,dim2b,dim2c}.

In our opinion, the $z$-dependent effective mass~\eqref{ansatz} should be viewed as primary while the dilaton background
as secondary, {\it i.e.} the standard SW holographic model follows after appropriate redefinition of fields.
If this viewpoint is correct, a question appears concerning the physical origin of $z$-dependence from a more fundamental
theory. Its origin might be inevitable when passing to non-conformal theories in the gauge/gravity correspondence.
We remind the reader that a 4D physical state in a dual 10D theory emerges in the form
\be
\Phi=e^{iqx}\psi(z,\Omega),
\ee
where $\Omega$ are coordinates on the 5D transverse space~\cite{pol} which usually represents sphere $S_5$.
In conformal case of pure $AdS_5$ space, a further factorization takes place,
\be
\label{fact}
\psi(z,\Omega)=Cv(z)g(\Omega),
\ee
where $g(\Omega)$ is a normalized harmonic in the angular directions which can be safely integrated out.
In non-conformal case, the factorization~\eqref{fact} is valid only at small $z$
because the infrared dynamics at large $z$ will in general induce mixing between different
harmonics~\cite{pol}. Nevertheless in AdS/QCD models, the factorization~\eqref{fact} is tacitly assumed for all $z$.
The effective $z$-dependence of mass term might originate from this mixing after integration over $\Omega$.
A holographic derivation of the form of this contribution 
remains of course an open problem.

Finally we would remark that the considered generalized SW model could find interesting applications in
calculating various transition matrix elements (hadron form factors) and the gluon parton densities
(parton distribution functions, transverse momentum distributions, and generalized parton distributions).
These quantities are defined as certain overlap integrals of normalizable modes with boundary currents which
propagate in AdS space~\cite{br}. It is known that calculations of these important physical quantities are
usually rather successful in soft-wall AdS/QCD (see, {\it e.g.}, a very recent work~\cite{Lyubovitskij}).
But we would expect a better quantitative agreement with the experimental data in case of considered generalized SW model
due to to its flexibility in fine tuning the actual spectrum of mesons via the intercept parameter. Also our
approach can be considered within the soft-wall Light-Front Holographic QCD~\cite{br} and help in solving the following problem:
The poles of vector bulk-to-boundary propagator in time-like domain ($q_n^2\sim n+1$) do not coincide with vector meson masses extracted from the bound-state
equation ($m_n^2\sim n+1/2$) and one needs to shift the pole positions manually to their physical location to obtain a meaningful comparison with measurements~\cite{br}.

\section{Conclusions}

The main result of our work consists in an explicit demonstration of the fact that
the AdS/QCD predictions for radial Regge spectrum of spin-1 mesons following from expansion of correlators at
high momentum can be reproduced from expansion at low momentum. We provided a set
of physical assumptions for which the whole scheme works successfully. Our analysis
places on a new quantitative level the general idea that bottom-up holographic models
interpolate between high and low energy sectors of QCD.

It is important to emphasize
that according to the ideas of gauge/gravity duality, the semiclassical holographic
dual models should describe the low-energy domain where QCD is strongly coupled. This
entails a much better conceptual justification for low-energy holographic predictions
in comparison with the usual high-energy ones. The fact that such low-energy predictions
can describe the whole radial spectrum looks encouraging.

Our approach can be extended to vector mesons interpolated by higher twist operators.
This includes consideration of non-zero constant $a$ in the ansatz~\eqref{ansatz}.
In the case of arbitrary $a$, the structure of 2-point correlators becomes different and deserves a separate study.
An extension to the scalar and tensor cases is straightforward. The development of
corresponding phenomenology is in progress.

\section*{Acknowledgments}

The work of S.S. Afonin is supported by the Russian Science Foundation, grant 21-12-00020 
(the results of Section 2 - the high energy holographic constraints). The work of T.D. Solomko
(the results of Section 3 - the low energy holographic predictions) was funded by RFBR grant 
for aspirants 19-32-90053.

\end{document}